# Streamer Blowout Coronal Mass Ejections: Their Properties and Relation to the Coronal Magnetic Field Structure

Angelos Vourlidas[1,2] and David F. Webb[3]
[1] *The Johns Hopkins University Applied Physics Laboratory, Laurel, MD 20723, USA*
[2] *Also at IAASARS, Observatory of Athens, Penteli, Greece*
[3] *ISR, Boston College, Chestnut Hill, MA 02459 USA*

### ABSTRACT
We present a comprehensive analysis of a particular class of coronal mass ejection (CME) event, called streamer-blowout CME (SBOs). The events are characterized by a gradual swelling of the overlying streamer, lasting hours to days, followed by a slow, wide CME, generally exhibiting a 3-part structure, which leaves the streamer significantly depleted in its wake. We identify 909 SBO events in the LASCO/C2 observations between 1996 and 2015. The average blowout lasts for 40.5 hours but the evacuation can take days for some events. SBO-CMEs are wider and more massive than the average CME. Their properties generally vary during and between solar cycles. Their minimum (maximum) monthly occurrence rate of one (six) events in cycle 23 has doubled in cycle 24---a probable manifestation of the weaker global fields in the current cycle. The locations of SBOs follow the tilt of the global dipole (but not from 2014 onwards), do not correlate with sunspot numbers and exhibit flux rope morphology at a much higher rate (61%) than regular CMEs (40%). We propose that these characteristics are consistent with SBOs arising from extended polarity inversion lines outside active regions (e.g. quiet sun and polar crown filaments) through the release, via reconnection, of magnetic energy, likely accumulated via differential rotation.

## 1. Introduction

Eclipses reveal the spectacular ray-like nature of the solar corona. The rays, called streamers, are tracers of the global magnetic field configuration of the Sun as their enhanced density marks the interface between opposite polarity open flux systems. Coronal Mass Ejections (CMEs) represent the explosive release of large amounts of magnetized plasma from closed magnetic field regions of the corona. The ejection process is thought to interact and disrupt the large scale magnetic field from active region to global scales. The most dramatic manifestation of this process is the disruption or, even destruction, of a streamer during a CME. In fact, the very discovery of CMEs was sparked by the disappearance of a streamer in successive orbits of the OSO-7 coronagraph (Tousey et al. 1972).

Named 'streamer-blowout' CMEs (SBOs, hereafter) by Sheeley et al. (1982), they are easy to recognize in coronagraph observations because they exhibit very particular signatures: the overlying streamer gradually swells over a period of a few hours to a few days, then a usually bright, well-structured, generally slow CME emerges from the streamer and leaves behind a depleted corona (Illing & Hundhausen 1986). This event sequence produces a characteristic 'bugle' signature in Carrington maps of the coronal brightness within a few solar radii ($R_s$) from the solar limb (Hundhausen 1993; Subramanian et al. 1999; see also Figure 1 in this manuscript).

Despite their ease of detection, there have been only three statistical studies of SBOs. Howard et al. (1985, 1986) computed some properties of these events in their analysis of *Solwind* observations between 1979-1984 (Table 1). They found that SBOs were slower but more massive than the full CME sample and that their occurrence rate appeared constant through the solar cycle. The third study specific to SBOs is Vourlidas et al. (2002) where they reported statistics for 147 SBOs in Large Angle and Spectroscopic Coronagraph (LASCO; Brueckner et al. 1995) C2 observations, on board the *Solar and Heliospheric Observatory* (*SOHO*) mission, from January 1996 to August 2001. Although in general agreement with the *Solwind* results, the study identified faster SBOs (with correspondingly shorter streamer swelling times) and indicated a solar cycle dependence on the occurrence rate on these CMEs.

As the increased sensitivity and cadence of the LASCO observations enabled Vourlidas et al. to extend the SBO definition to faster and relatively smaller streamer blowouts, so it allowed Sheeley and coworkers to uncover the fainter end of the streamer blowout spectrum via a synthesis of past observations of inflow/outflow pairs (in/out pairs) and other faint dynamic phenomena. First, Wang & Sheeley (2006) argued that in/out pairs and arch-like streamer disruptions were signatures of small-scale flux rope formation and detachment from different viewing angles. Then, Sheeley & Wang (2007) showed that 80% of these in/out flow pairs were associated with protracted (1-2 days), slow (~20 km/sec) streamer inflation and eventual streamer detachment. Outflow-only events (the 'arch-like' features) were likely SBOs seen face-on (see their Figure 17 for an example). Sheeley et al. (2007) argued that all these observations can be brought into a common picture by considering the height where magnetic reconnection sets in to detach the flux rope and streamer into an SBO and projection effects. If reconnection occurs in the coronagraph field of view (say, at ~4 $R_s$) then an in/out pair and flux rope may be seen. If it occurs in the inner corona or it is occulted because the event is away from the sky plane, then only an outward component appears. CMEs starting at lower heights are generally brighter ('classical' SBOs) unless they propagate away from the sky plane, thus appearing as arch-like structures. In other words, arch-like fronts, 'open-jaw' signatures (i.e., Figure 15 in Wang & Sheeley 2006), and small-scale V- (or Y-) shape disconnection events (i.e., Figure 3 in Wang & Sheeley 2006) are different projections or substructures of SBOs.

The driver of these events is not yet settled. Sheeley et al. (2007) argue for a non-magnetic driver. They suggest that SBOs do not follow the sunspot cycle, are the dominant events during solar minimum, and hence originate from changes (due to coronal heating) in the balance between magnetic tension and the plasma pressure gradient. In that case, they should occur away from active regions and dominate during solar activity minima. They should still entrain a flux rope as all CME models require (Chen 2011) since the final release of the CME material will be due to magnetic reconnection. These predictions could be verified with a thorough analysis of the properties of these events.

To describe SBOs and answer their origin questions, we compile a comprehensive set of statistics of SBO properties in LASCO observations. We extend the Vourlidas et al. (2002) work to the end of 2015, thus covering 19 years of coronagraph observations over two solar cycles. We compare this catalog with our recently completed post-CME ray catalog (Webb & Vourlidas 2016) and with surface signatures to examine the origin of these events. We also discuss the role of SBOs in large-scale field evolution.

The paper is organized as follows. In Section 2, we describe our assumptions for the event selection, the resulting database and statistical measurements. In Section 3, we analyze the event statistics and compare them with surface activity, (pseudo)streamer structure and post-CME rays. We discuss our findings in Section 4 and conclude in Section 5.

## 2. SBO Selection and Measurements

The original SBO-list in Vourlidas et al. (2002) covered the period from January 1996 to December 2001. We extend the list to December 2015, thus covering the whole of Solar Cycle 23 (SC 23) and past the maximum of SC 24. In the 2002 paper, we identified the SBO events based on their 'bugle'-like appearance at 3 Rs in synoptic coronal maps. In the intervening years, we continued to compile a list of interesting events (including SBOs) by visual inspection of the direct and running difference monthly C2 movies, available on the web[1]. This list forms the basis for the SBO catalog here.

### 2.1 SBOs in Carrington Maps

As mentioned earlier, the brightness evolution of SBO-CMEs typically produces the characteristic signature of a 'bugle' in Carrington maps of coronagraph data. The maps are constructed by taking a daily 180° slice, with 1° resolution, at a given heliocentric radius over either the East or West limbs and stacking those slices in time for a full Carrington Rotation. These maps have been available online[2] since the beginning of the *SOHO* mission and extend to Carrington rotation 2168 (Sep-Oct 2015) as of this writing. They provide a concise visualization of the evolution of the coronal structure, in white light, and have been used for CME detection and measurements (e.g., Lamy et al. 2017, and references therein).

In Figure 1, we provide a few examples of SBOs in Carrington maps. The top panel shows one of the longest blowouts in our sample, occurring over October 10-17, 2003 (dashed arrow). The event is very bright, occupying over 40° in position angle at its peak. The streamer depletion is complete. In Carrington Rotation 2045 (bottom panel), we have two SBOs with clear flux rope morphology. In the maps, both CMEs exhibit a 'bean'-like shape, which is likely a result of their very circular appearance in the direct images. Note that a new streamer appears at the northern end of both CMEs. The second event (on November 24, 2003) appears to mark the end of a long-lived streamer.

---

[1] http://lasco-www.nrl.navy.mil/daily_mvi/

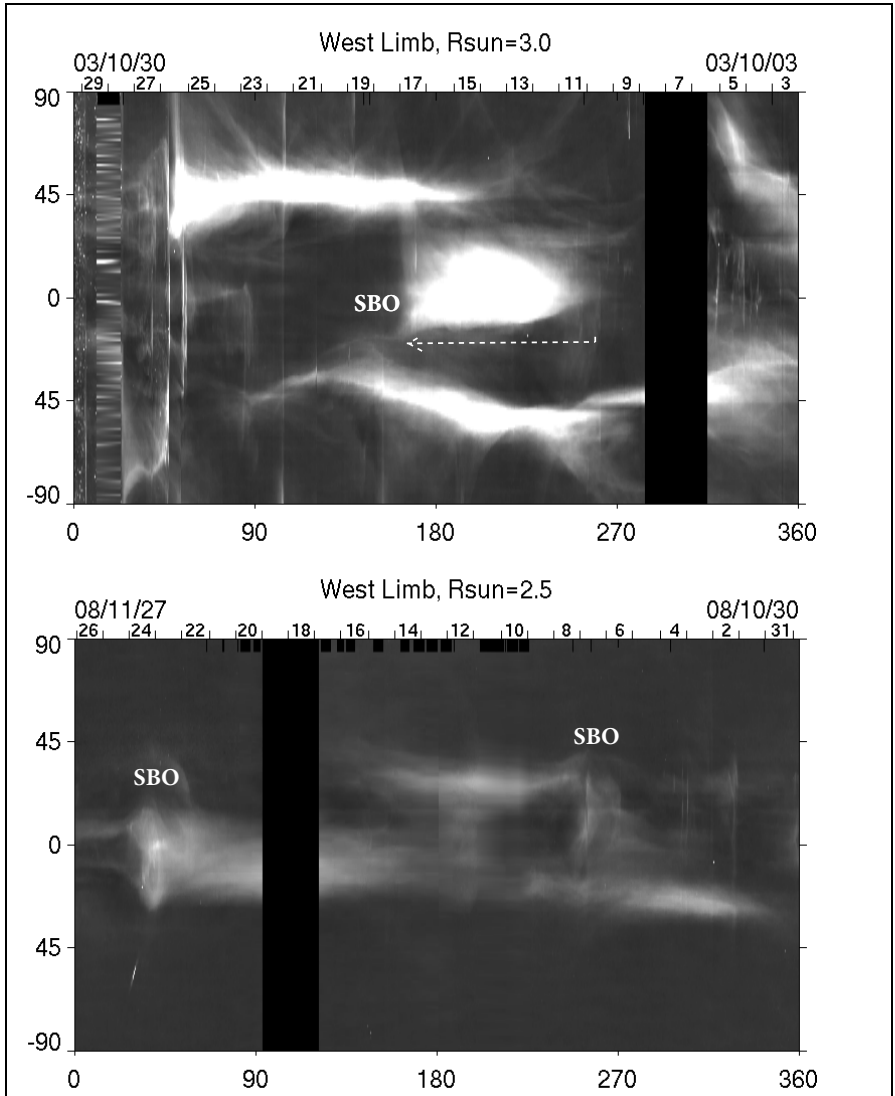

Figure 1. Examples of SBO-CMEs in LASCO Carrington Maps. Time runs from right to left. Top panel: An exceptionally long-duration SBO (~4 days) during the SC 24 maximum (dashed arrow). The event shows the characteristic 'bugle' morphology. Bottom panel: Two SBOs during SC 23/24 minimum exhibiting flux rope morphology, represented by the bean-like shape in this projection. These Carrington maps were built by LASCO-C2 observations at 3 (top) and 2.5 Rs (bottom) over the west limb and are available online[2]. The dark columns denote missing data.

## 2.2 Selection Criteria

The criteria for categorizing a CME as an SBO follow the original definition for such an event given in Howard et al. (1985). In particular, we scan the monthly C2 movies to search for evidence for streamer swelling followed by an eruption and subsequent evacuation of that streamer. We use only the LASCO C2 images for the event identification because streamers and their cusps are best seen below 4-5 Rs.

---

[2] https://lasco-www.nrl.navy.mil/carr_maps

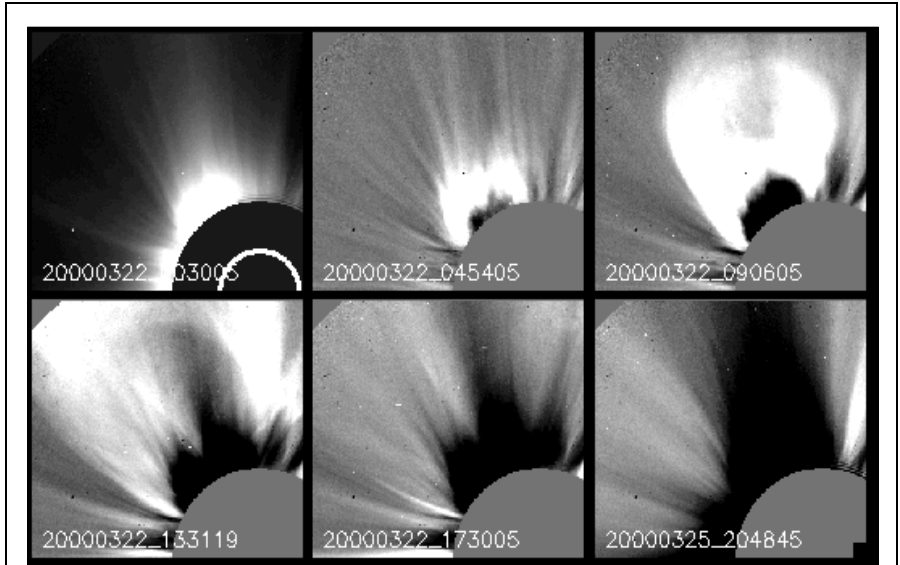

Figure 2. Example of SBO on March 22, 2000 in LASCO/C2. The image at 00:30 UT is the pre-event image and all subsequent images are differences from that image. The CME is out of the C2 field of view by 17:30 UT but the coronal evacuation continues until about 20:48 UT on March 25, making this an approximately 3-day event.

We put no time limits on the duration of the swelling as long as it is longer than ~ 1 hour, corresponding to 3-6 C2 image frames (the C2 synoptic cadence of 24 min increased to 12 min since August 2010). One hour is about the shortest time to identify a streamer motion, based on our experience. The start of the streamer swelling is generally hard to define because of its gradual nature. Running difference images show very little change, requiring the examination of long time series and frequent 'back-and-forth' frame switching. The swelling is detected via the appearance of excess brightness enhancement across the top of the streamer (Figure 2, see also Illing & Hundhausen 1986). In several cases, we had to employ base-difference images to detect the start of the swelling. The exact timing of the swelling is hard to define visually and to separate from the effect of solar rotation, so we estimate the error in the start time to be again ~3 images (or 1 hour prior to 2010 and 36 mins since 2010). Consequently, the end of the event occurs when the streamer reaches minimum brightness. Again, this is a subjective criterion. In many cases, the streamer does not disappear completely (partial-SBO) or another streamer stalk appears while material from the SBO is still outflowing. In those cases, we set the end of the event at the time of the last visible outflow.

For the properties of the CME itself, i.e. speeds, width, etc., we rely on the LASCO CDAW list measurements for that particular event. Because we derive our SBO list independently, we have to search the CDAW list for the event corresponding to our SBO. We do this semi-automatically, based on the position angle and the time of first image in C2. If we find more than one CDAW entry, then we select the entry corresponding to our event by manually inspecting the CDAW candidates. Because some events may be misidentified with this method, we made a detailed manual check of 40 events across the database and found that only one of them had incorrect CDAW parameters. Therefore, we estimate that 1-2% or 11 events in the whole list may have the wrong identifications. This percentage is small enough to not affect the statistical analysis presented here.

## 2.3 General Properties of SBOs

Our final list contains 909 SBOs observed from 1996 – 2015. We were able to cross-link 869 of them with the CDAW list. Of those, 835 have speed measurements, 785 have mass measurements, and 783 have kinetic energy entries. The statistics for these events are compared to the overall LASCO CME sample for the 1996-2015 period, as well as the *Solwind* results in Table 1. To make comparisons more robust, we report statistics only for events with 10°-180° widths, and exclude events with no speed or mass measurements. The same criteria are used for the CDAW list. This is the reason for varying sample sizes in the SBO column. We do not have any such information for the Solwind list, so we present only the reported statistics from Howard et al. (1985)..

The LASCO SBOs are, on average, wider, faster, and have higher kinetic energy, but lower mass, than the much smaller *Solwind* sample. The slower speed and higher average mass of the *Solwind* sample is likely an effect of the lower duty cycle and sensitivity of that telescope. Slower, more massive events would be more readily detected. The higher speed in our sample is driven somewhat by the cadence of C2 observations but also by our allowance for short streamer swellings of only a few hours. Streamer variability at such short scales would be difficult to discern with the *Solwind* duty cycle.

*Table 1. Comparison of SBO average parameters with their standard deviation to the full CME sample for 1996 - 2015 and past SBO measurements. The number of events with available measurements in each quantity are shown in the parentheses. Only events with widths 10°-180° are considered.*

| Parameter | LASCO SBOs (909 events) | *Solwind* SBOs (50 events) | All LASCO CMEs (22606 events) |
|---|---|---|---|
| **Width (Deg)** | 72 ± 33 (847) | 44 | 49 ± 35 |
| **Speed (km/s)** | 390 ± 233 (813) | 200 | 365 ± 209 |
| **Mass ($10^{15}$ gr)** | 2.9 ± 3.3 (762) | 5.4 | 0.9 ± 3.4 |
| **$E_{kin}$ ($10^{30}$ ergs)** | 3.3 ± 21 (762) | 0.56 | 1.5 ± 10 |
| **Duration (hrs)** | 40.5 (909) | Not measured | Not measured |
| **Reference** | This work | Howard et al. (1985) | This work |

Compared to the overall CME sample, the LASCO SBOs are wider and more massive than the average CME. The full CME statistics are biased downwards by the inclusion of small events, such as jets, but the approximate factor of 3 difference in the average mass is a direct reflection of the nature of SBOs. These events remove a lot of mass from the corona.

We find that SBOs are long-lived events. The average event lasts 40.5 hours (the median is 35.4 hours) but some can go on for several days (Figure 3). For comparison, an CME moving at 400 km/sec will take 13.4 hours to cross the combined LASCO C2-C3 fields of view. Most events (histogram peak in Figure 3) last for 25.9 hours. Isolating the duration of the streamer swelling itself is challenging because it is hard to discern when the swelling stops and when the CME starts. For many of the events, for example, the time of the first appearance of the CME in C2 (used as the CME time in the CDAW catalog), coincides or nearly coincides with the start of the swelling in our measurements.

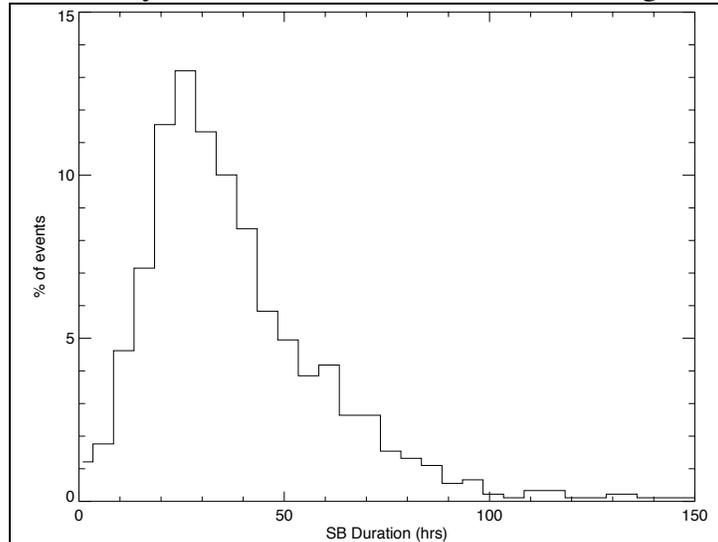

Figure 3. Distribution of event durations, in hours, for the LASCO SBOs in 1996 to 2015 (909 events). A typical SBO lasts for about one day (25.9 hours).

Since the event duration is one of the characteristics that set these events apart from the rest of the CMEs (the other is the disappearance of the streamer), we examine the variation of some key parameters as a function of event duration in Figure 4.

For speed, we use the final speed at the edge of the LASCO-C3 field, based on a 2$^{nd}$ degree fit. There is considerable spread in the speeds, as expected given projection effects (Figure 4, top left). To reveal any trends, we bin the speeds in 10-hr duration bins, extract their statistics (mean and standard deviation) and overplot them in the figure. The average speed shows a trend towards lower speeds for longer duration events, suggesting that the latter may be less energetic events. There is a hint of a plateau in the speeds, at around 300 km/s, which can be easily understood because any ejection needs to attain a speed close to the ambient solar wind in order to be expelled into the heliosphere. The mass plot (top right panel) shows a similar trend with less massive events as the duration increases. The kinetic energy (bottom left) makes this trend clearer (being the combination of speed and mass).

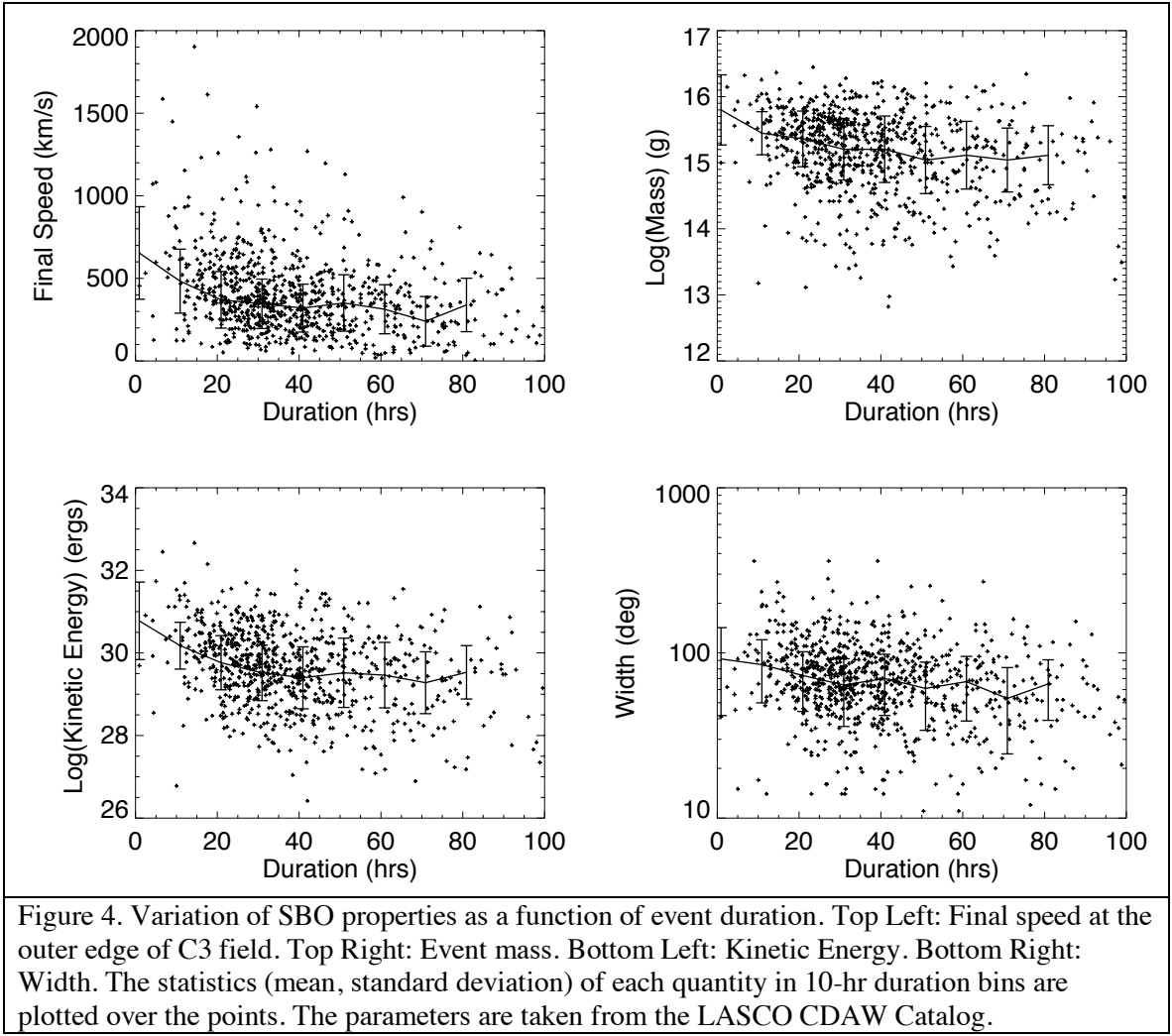

Figure 4. Variation of SBO properties as a function of event duration. Top Left: Final speed at the outer edge of C3 field. Top Right: Event mass. Bottom Left: Kinetic Energy. Bottom Right: Width. The statistics (mean, standard deviation) of each quantity in 10-hr duration bins are plotted over the points. The parameters are taken from the LASCO CDAW Catalog.

Finally, the widths of SBOs follow the general trend and suggest that long duration events are narrower. In other words, the variations of the quantities in Figure 4 suggest that slower, smaller, less massive and energetic events take longer to evacuate the streamer, resulting in blowout events that can last for days.

### 3. Analysis of the SBO-CME Properties

For this paper, we restrict our analysis to the general properties of the SBOs, such as their solar cycle variation, occurrence rates, dynamic properties such as duration and energies, and their associations with other phenomena, such as prominence eruptions and post-CME rays (e.g., Webb & Vourlidas 2016).

## 3.1 Solar Cycle Variations

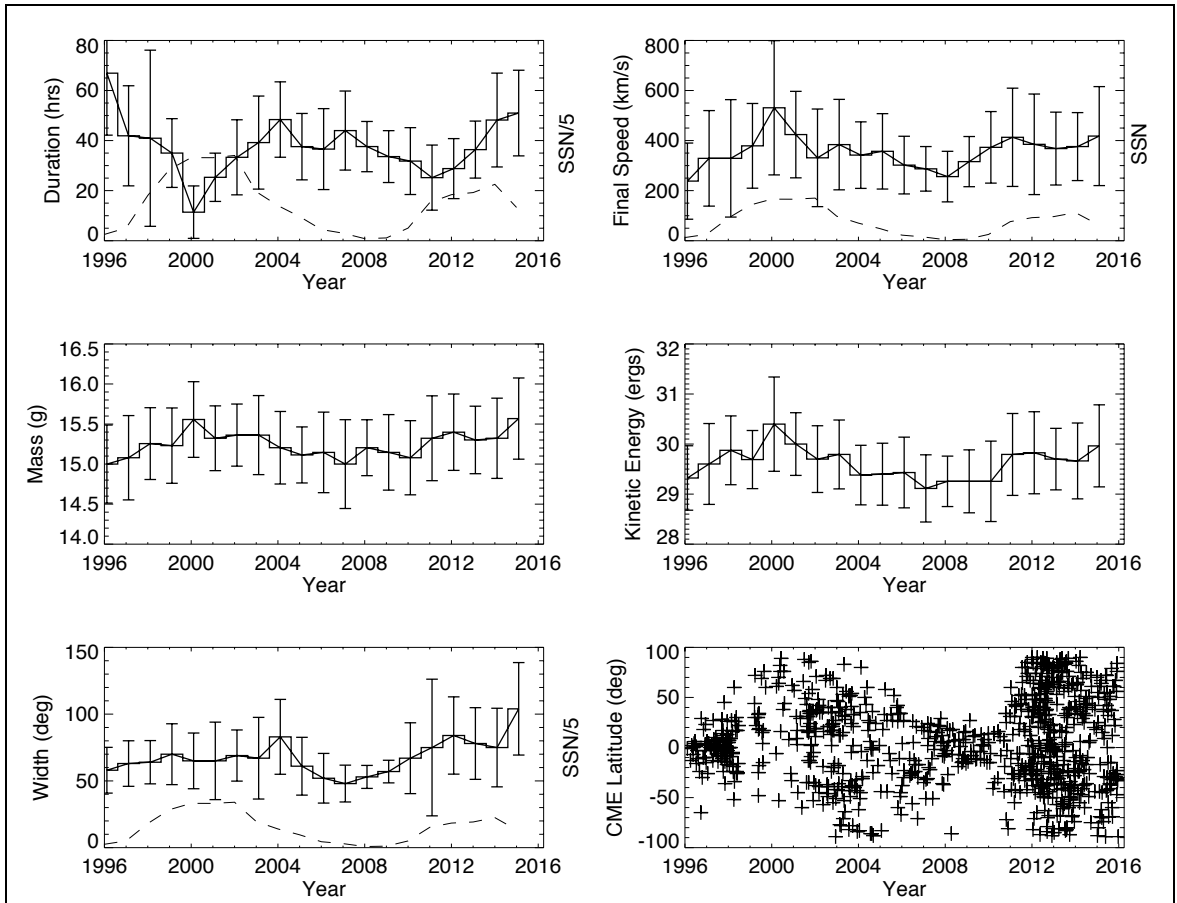

Figure 5 Yearly averages of SBO-CME properties from 1996 - 2015. Top Left: SBO duration. The bars denote the range in each parameter for the given bin. Top Right: SBO Speed at the edge of C3 FOV. Middle Left: SBO Mass. Middle Right: Kinetic Energy. Bottom Left: Width. Bottom Right: Central Position Angle. The dashed lines show the yearly average of the sunspot number (divided by 5, except in the speed panel). All parameters, except event duration, are taken from the LASCO CDAW Catalog.

We plot yearly averages of several SBO parameters in Figure 4 to examine their solar cycle dependence. The bars are the standard deviation of a given parameter calculated within the yearly bin and reflect the variability, not the error, in the measurements of that quantity. The top left panel shows the event duration as a function of time. The most striking feature is the sharp reduction in the SBO duration as SC 23 rises to its first maximum. The average event duration is reduced from about 66 hours, in 1996, to only 10 hours, in 2000 and then rises as the cycle wanes. However, the durations never reach the high values measured during the SC 22/23 minimum. Instead, they hover at around 40 hours through SC 24, with a small dip in 2011 reminiscent of the decrease in 2000. The duration increases with a slope similar to the post-SC 23 maximum all the way to the end of the available measurements. We believe that the difference in the duration behavior between the two cycles may reflect the difference in the dipole tilt between SC 22/23 and SC 23/24 minima. We expand on this argument at the end of the section.

The final speed (top right panel) increases from 250 to 560 km/s between 1996 and 2000 and gradually returns to 250 km/s in 2008, at the SC 23/24 minimum. It then rises to about 410 km/s by 2011, following the SC 24 rise towards maximum but it remains flat through the rest of the available SC 24 measurements, in contrast to the SC 23 behavior. The speeds have the opposite trend to the event duration, as expected, but they do not show a difference between the SC 22/23 and SC 23/24 minima, nor do they show the two-peaked pattern for the SC 23 maximum, as many other solar parameters do, including the statistics of the full CME sample (e.g. Vourlidas et al. (2010), Figure 14). In other words, the SBOs do not seem to correlate with sunspot numbers and consequently with the evolution of strong magnetic fields. A possible explanation is that SBOs originate from PILs outside of active regions.

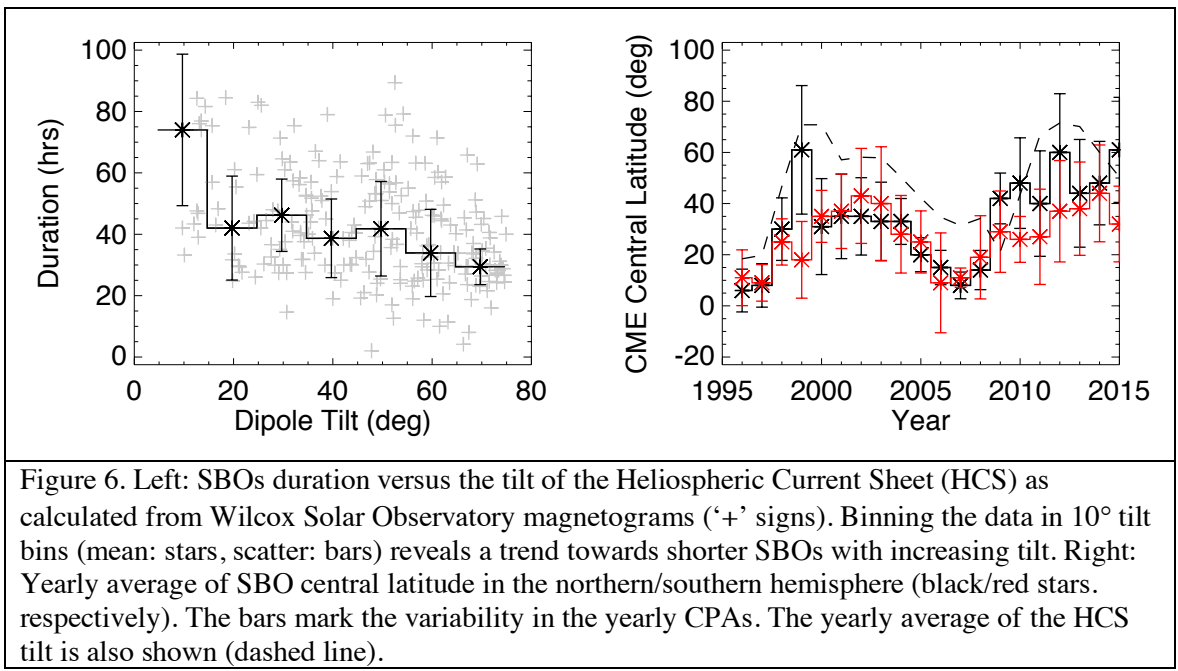

Figure 6. Left: SBOs duration versus the tilt of the Heliospheric Current Sheet (HCS) as calculated from Wilcox Solar Observatory magnetograms ('+' signs). Binning the data in 10° tilt bins (mean: stars, scatter: bars) reveals a trend towards shorter SBOs with increasing tilt. Right: Yearly average of SBO central latitude in the northern/southern hemisphere (black/red stars. respectively). The bars mark the variability in the yearly CPAs. The yearly average of the HCS tilt is also shown (dashed line).

The average mass (middle left panel) shows little variation with solar cycle, with some increase from minimum to maximum. Of course, there is variation in individual events, as some reach about $7 \times 10^{15}$ g, mostly during high activity periods (e.g., 2000 and 2003). This is not apparent in SC 24. The mass increases in 2011, remains more or less constant through the SC 24 maximum (2012-2014) but then increases again in 2015 instead of decreasing as happened in the SC 23 decline. The mass increase is likely due to the increase in CME widths seen in the bottom left panel.

The average kinetic energy (middle right panel) follows the evolution of the speeds but shows a minimum in 2007 rather than 2008 (speed minimum). There is a sharp increase in 2010 which coincides with the onset of SC24 activity and is seen in some of the statistics of the full CME sample (Vourlidas et al. 2017). However, the SC 24 kinetic energy never reaches the SC 23 levels, although there is an upward trend in 2014-2015.

The average width (bottom left panel) of SBOs shows some interesting patterns. It remains relatively constant, between ~60° and 70°, throughout the rise and peak of SC 23, then narrows to about 50° at the SC 23/24 minimum and rises almost monotonically to 100° by 2015. The width peaks, not at the SC 23 sunspot maximum but at the beginning of the declining phase in 2004. The behavior is very different in SC 24. Instead of a reduction in the SBO widths as the cycle starts to decline, we see a rather sharp increase, with 2013 being the apparent turning point. To make sure that this difference does not arise from the way we measured the events between the 2000s and now, we reprocessed nine random months distributed between 2001 and 2007 associating manually the events to the CDAW measurements as we did for the 2012-2015 events. We found no differences from our original list, therefore, the change in the SBO width is a reflection of the wider nature of SBOs in SC 24 compared to SC 23 and may be a result of the weaker open fields in SC 24 as suggested by Gopalswamy et al. (2014).

Finally, we plot, in the bottom right panel, the central position angle (CPA) of these events. The scatterplot shows that the CPAs of the SBOs follow the expansion of the streamer belt, from its narrow equatorial location at the SC 22/23 minimum to its spread around the Sun at SC 23 maximum and its return to a nearly flat belt in SC 23/24 minimum. Between 2009-2010, the CPA sharply spreads to all latitudes. Again, this is similar to the behavior in the speeds and kinetic energies and signifies an intrinsic change in the large-scale magnetic field that marks the onset of SC 24. However, in sharp contrast to SC 23, there is no evidence that the CME CPAs retreat to lower latitudes during the declining phase of SC 24.

We propose that the cycle evolution of both the event duration and CPA can be understood as the effect of the solar dipole tilt, at least partially. We expect longer durations for SBO events during solar minima when the HCS aligns with the solar rotation axis, resulting in a flat and bright streamer belt. In that case, a streamer blowout would require significant mass loss over a wide longitude to be visible in the images. Given the slow speeds of solar minimum events (Figure 5), the durations of the SBOs would be longer. In contrast, at solar maximum, the HCS is highly tilted towards the equator, revealing its undulations (streamers, in other words) and presenting a smaller line of sight to the observer. Consequently, it takes less 'effort' to create a blowout in the images. We do not necessarily expect higher mass for solar minimum SBOs, however. The final mass depends both on the line-of-sight (longitudinal) extent as well as on the coronal density, which tends to be higher during solar maximum. The latter effect seem to be more important according the Figure 5 mass statistics that show higher SBO masses at towards solar maximum.

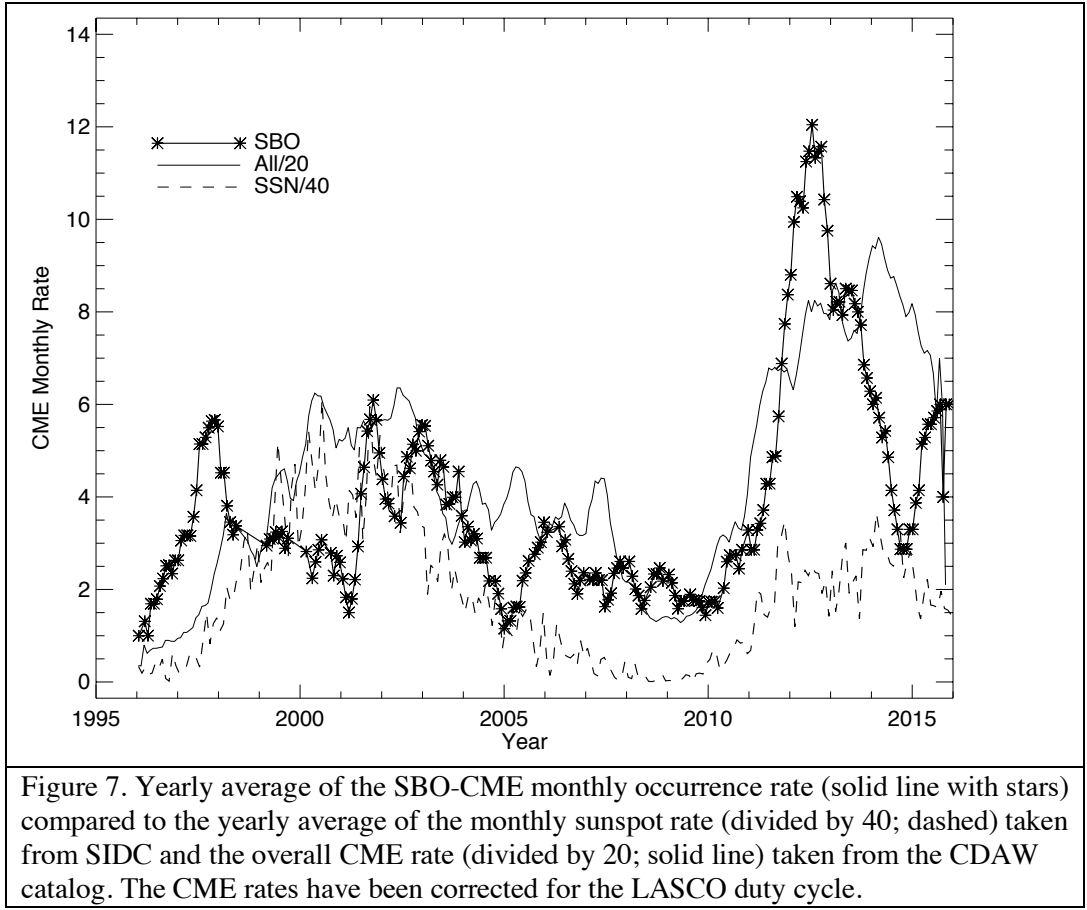

Figure 7. Yearly average of the SBO-CME monthly occurrence rate (solid line with stars) compared to the yearly average of the monthly sunspot rate (divided by 40; dashed) taken from SIDC and the overall CME rate (divided by 20; solid line) taken from the CDAW catalog. The CME rates have been corrected for the LASCO duty cycle.

To test this, we obtain the 'tilt angle' of the Heliospheric Current Sheet (HCS) from the Wilcox Solar Observatory (WSO)[3] and use the classical average value, $L_{av}$ based on the line-of-sight magnetograms. We then examine the relation of the event duration and position angle to the tilt as a function of the solar cycle (Figure 6). Again, the scatterplot of duration versus tilt (Figure 6, left) shows some scatter, most likely arising from projection effects (as does most of the scatter in CME quantities, e.g. Vourlidas et al. 2010). If we bin the duration in 10° tilt bins, the trend becomes clearer. Longer durations are associated with small inclination of the dipole that occurs during solar minima. The right panel of Figure 6 reinforces this conclusion. Here, we plot the yearly average and standard deviation of the SBO-CME central latitude in each hemisphere (derived from the event CPA). The yearly average of the dipole tilt is overplotted with the dashed line. The correspondence between the two quantities is obvious (see also the scatterplots of the CME latitude values in the bottom right panel of Figure 5), leading us to conclude that: (1) the SBO-CME locations follow the HCS tilt, and (2) the observed SBO durations may be driven by the thickness of the plasma sheet along the line of sight or equivalently by the orientation of the HCS. However, this relation appears to break down after 2014 when the inclination of the dipole tilt is getting smaller but the SBO CPAs remain constant. This change in behavior may be another manifestation of the unusual SC 24 seen in other physical parameters (e.g. Zerbo & Richardson 2015). We have no explanation to offer at this moment. The accumulation of more SBO-CME statistics over the next couple years may shed some light on this behavior.

---

[3] http://wso.stanford.edu/Tilts.html

**3.2 Occurrence Rates**

We turn our attention to the occurrence rate of SBO events. In Figure 6, we plot monthly averages of the event rate over a Carrington rotation, instead of a daily rate, because of the relatively small number of SBO-CMEs compared to the total CME numbers. The rates have been corrected for the LASCO duty cycle based on the detection of a 500 km/s CME (see Vourlidas et al., 2017, for details). Their occurrence rate shows no clear relation to the sunspot number (dashed line) in SC 23. Instead, we find a peak associated with the onset of SC 23 activity (in 1998) and two more peaks in 2002 and 2003, associated with the second sunspot peak of SC 23. The SBO minimum occurs in 2005, rises in 2006, and remains more or less constant at ~ 2 SBOs/month throughout the SC 23/24 sunspot minimum. This is double the SBO rate during the SC 22/23 minimum (~1 SBO/month). The equivalent overall LASCO CME rate during these minima was one CME every 1.5-2 days (Webb & Vourlidas, 2017). For SC 24, there is a notable sharp rise in the number of SBOs from 2010 to 2012, broadly similar to the SC 23 rise, but reaching a much higher rate of 12 SBOs/month), compared to 6 SBOs/month in SC 23.

In summary, the SBO occurrence rate varies with the solar cycle but not in tandem to the sunspot number or the overall CME rate. It seems to follow the increase in sunspot numbers during the rise phase of the Cycle but does not follow the sunspot evolution during the maximum phase. The SSN is a measure of the strongest magnetic fields, but it is well correlated with the global photospheric magnetic flux. The behavior during the declining phase of SC 24 remains to be seen as we have only the SC 23 statistics so far.

Overall, the interpretation of Figures 6-7 favors a magnetic origin for SBOs but not from active regions given the lack of correlation with sunspot numbers. Streamer blowouts could originate from extended neutral lines arising from dispersed flux systems outside active regions, such as polar crown filaments. We explore this idea later in the manuscript (sections 3.5-3.6 and discussion).

## 3.3 Morphology

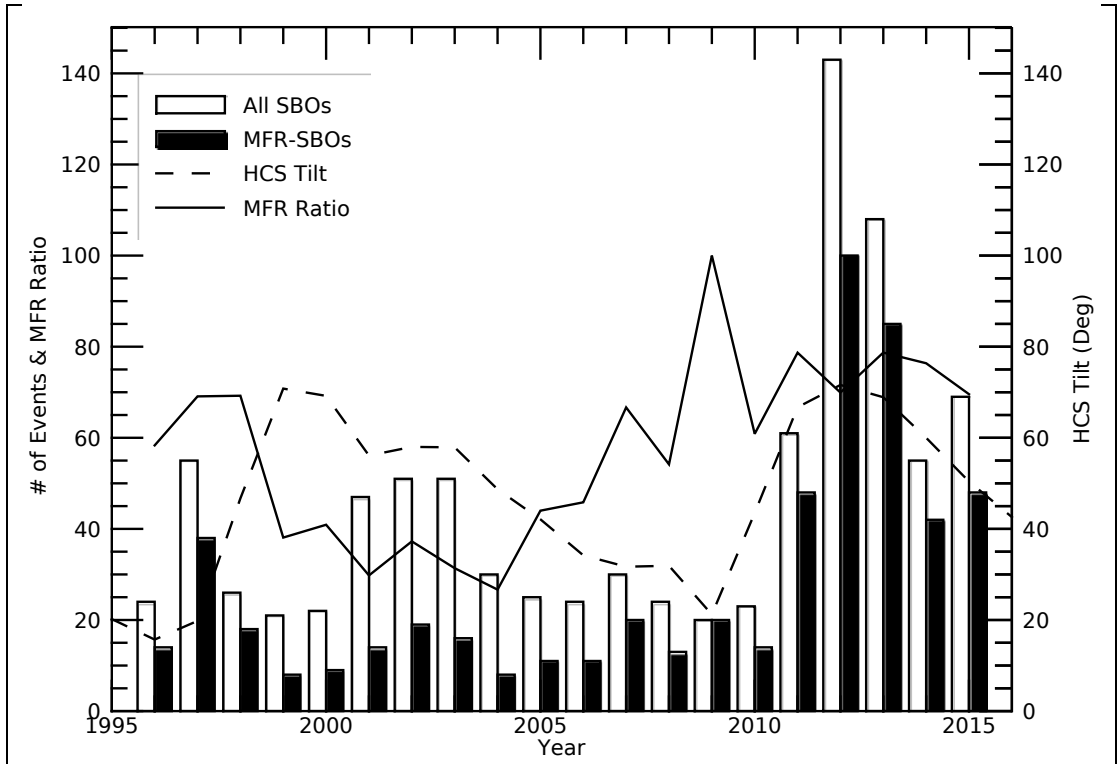

Figure 8. The yearly number of SBO-CMEs with flux rope morphology compared to the total number of SBOs in our sample (bars) and the ratio of MFRs to total SBOs (solid line). The dashed line corresponds to the HCS tilt, computed from WSO magnetograms. The flux rope morphology becomes clearer when the HCS is aligned with the observer (low tilt) implying that the axes of the flux ropes are generally aligned to the HCS for SBO-CMEs. This trend holds through the SC23/24 minimum but not for SC24.

SBO-CMEs, often exhibit well-defined internal structure and tend to be labeled as '3-part' CMEs. As we argued recently (Vourlidas et al. 2013; 2017), the internal structure is consistent with that of a magnetic flux rope (MFR). During the buildup of the catalog, we identified those events with MFR morphology using the criteria in Vourlidas et al. (2013). We find that the MFR morphology is very common in SBOs, with 61% (556/909 events) showing clear MFR signatures. This is a significantly higher percentage that the overall CME sample (40%; Vourlidas et al. 2013). The percentage of MFRs varies greatly with solar cycle, from 26% in 2001 to 100% in 2009 (Figure 78). Also, there appear to be more MFRs at the same cycle phase in SC 24 compared to SC 23. The MFR occurrence rate remains above 61% since 2009.

Since the apparent CME morphology heavily depends on projection (Vourlidas et al. 2017), we examine whether the HCS orientation may explain the solar cycle behavior in the percentage of MFRs by overplotting the HCS tilt in Figure 8 (dashed line). Indeed, the appearance of MFRs seems to follow the HCS orientation for SC 23. The percentage of flux rope morphology rises as the HCS tilts lie below about 40° and wanes as the HCS axis tilts towards the equator. Since the flux rope morphology is easier to detect when the MFR is seen along its axis because of the longer line-of-sight integration, the relation between HCS tilt and MFR percentage in Figure 8 suggests that the MFRs in these CMEs are closely aligned with the HCS.

In other words, SBOs, particularly during minimum, seem to be eruptions from large scale neutral lines ('streamer eruptions') rather than eruptions from active regions. The drop in MFR detections during maximum is consistent with this interpretation. During these periods, the large scale neutral line is highly distorted so the axis of the erupted flux rope, even if it is locally aligned with the large scale neutral line, may not have a favorable projection in the images, thus reducing its detectability. We cannot exclude the possibility that some of the SBOs during maximum originate from ARs. Strong magnetic regions tend to distort the HCS, creating local folds that appear as (generally narrower but brighter) streamers. Eruptions from such configurations will also result in a blowout and may, hence, be included in our statistics.

However, these arguments do not hold for SC 24. Although the percentage of MFR-SBOs drops to 70-76% during SC 24 (2012-2014), they are far more numerous than in SC 23 (as is the overall SBO rate). This yet another interesting aspect of the unusual SC 24 for which we have no immediate explanation.

In summary, we find that SBO-CMEs tend (1) to contain magnetic flux ropes and (2) to be aligned to the HCS.

**3.4 Association with post-CME rays**

We surveyed the association between SBOs and post-CME rays using our recently published catalog (Webb & Vourlidas 2016). In white light images narrow, coaxial rays trailing the outward-moving CME have been interpreted as current sheets, which in turn indicate the magnetic disconnection of the CME from the near-surface magnetic arcade. Since Webb & Vourlidas found that most of the CME-rays were associated with SBOs, it is of interest to compare the occurrences of those CME-rays with the full SBO catalog.

*Table 2. Occurrence of post-CME rays in SBOs*

| Year | SBOs with rays/ all SBOs | | All CMEs[1] |
|---|---|---|---|
| 1996 | | | |
| 1997 | 14 / 105 | 13.3 % | 11.4 % |
| 1998 | | | |
| 2001 | 4 / 47 | 8.5 % | 7.2 % |
| 2010 | 0/ 23 | 0 | -- |
| 2011 | 9/ 61 | 14.8 % | -- |

[1]**Webb & Vourlidas (2016)**

We did not perform a thorough investigation but rather checked the associations during a few full years covering the range of our SBO sample. This exercise (Table 2) confirms the Webb & Vourlidas result that the occurrence rate of SBO CMEs with rays is higher around solar minimum than maximum, likely related to the more complex coronal background at maximum. In addition, post-CME rays occur in SBOs at slightly higher rates at both maximum and minimum compared to the whole CME population. Whether this difference is significant requires a more thorough analysis of the events and their sources regions which is beyond the scope of this paper.

### 3.5 Association with PILs and prominences

What about the association of SBOs with surface features/activity, particularly active regions and sunspots or filaments/prominences and long quiescent PILs such as the polar crowns? Recently, Webb et al. (2017) demonstrated that a basal rate of CMEs exists during the last four solar activity minima. These CMEs appear mostly as gradual reconfigurations of the helmet streamer structures that characterize the flattened HCS. Many of these are SBO CMEs and these ejections are likely related to a minimum threshold for magnetic energy dissipation and/or the ejection of magnetic helicity from the Sun.

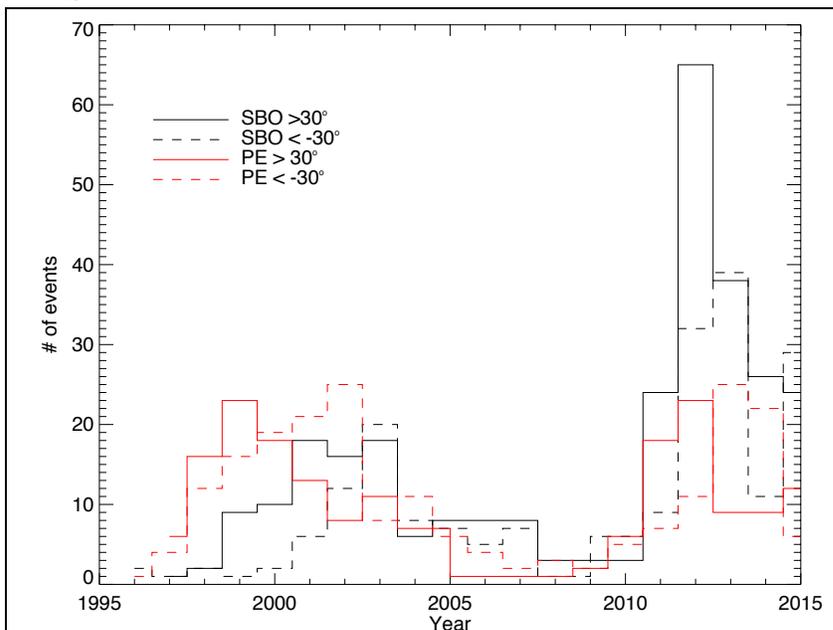

Figure 9. Yearly averages of SBO CMEs (black lines) and prominence eruptions (PE) detected with the Nobeyama Radioheliograph (red lines) outside the streamer belt (defined as position angles within 30° of solar equator).

The source regions of the streamers and their associated CMEs, including SBO CMEs at solar minimum appear to be along the global PIL that is the base of the HCS. Some of the streamer-disruption CMEs were associated with prominence eruptions (PEs), perhaps 20% of the minimum-CMEs, or about 2 per month. Independently we compared the SBO list in 2011 with a list of SDO/AIA PEs[4]. Taking those PEs listed as "Behind the Limb" or "At the Limb", we found that of the 60 SBOs in 2011 ~23 PEs were associated. Thus, this result shows that at least during the rise of SC 24 there is a reasonably good correlation between PEs and SBO-CMEs.

The Nobeyama Radioheliograph (NoRH), now operated by the Solar-Terrestrial Environment Laboratory at Nagoya University, began observing limb prominence eruptions at 17 and 34 GHz in 1992. Over the time period of the LASCO observations, we compared the NoRH list of PEs with our SBOs in the two hemispheres. Figure 9 shows this annual distribution of PEs and the CPAs of SBOs above 30° heliolatitude in both hemispheres where PEs are likely arising from polar crown filaments. The two phenomena occur in similar numbers (during the two solar minima) and follow similar trends (with the possible exception of the SC 23 rise phase). Even with the large number of SBOs during the rise of SC 24, the evolution of PEs and SBOs is generally similar. We assert that there is a strong relationship between SBOs and high latitude (i.e polar crown) filament eruptions. We plan to investigate this further by making one-to-one comparisons between the data sets.

---

[4] http://aia.cfa.harvard.edu/filament/

## 4. Discussion

We investigate the properties of a type of coronal ejection event undergoing a set of well-defined evolutionary steps in coronagraph images. The event starts with a slow *swelling* of a streamer that can lasts about a day, on average. Eventually, a (usually clear) *3-part CME* appears from behind the occulter and is released slowly into the outer corona, followed by *plasma outflow* that can also last for many hours, leaving behind a *depleted* streamer. The streamer usually reforms at a slightly different position angle, implying a *change* in the global magnetic field configuration as a result of the ejection.

None of these evolutionary steps is particularly unusual for CMEs at large, with the exception of the streamer depletion, which, in many cases, serves at the primary means of detection for these events. Hence, the label 'Streamer Blowout CMEs' is an appropriate description for these CMEs, even though the streamer may not deplete completely in all events. Sometimes, this may be due to projection effects (intervening structures) or due to the CME reconfiguring only a small portion of the streamer belt. In any case, the detection of streamer depletion, even in Carrington maps (Figure 1), is an indication that SBO-CMEs originate in extended polarity inversion lines.

Are there different classes of SBOs? We do not believe so. The statistics show no evidence for different SBO populations. The events have very similar appearances in the coronagraph images. Apart from the streamer swelling and subsequent depletion, SBO-CMEs look like the average CME. An obvious question is whether the streamer blowout phenomenon occurs in pseudostreamers as well. If it did, it would imply that the origins or drivers of SBOs may be non-magnetic (since pseudostreamers lack a current sheet). Wang (2015) studied a group of 8 CMEs arising from pseudo-streamers. We compared these with our SBO list and found that none of these 8 events were associated with SBOs. Therefore, this limited sample suggests that pseudo-streamers do not yield SBO CMEs.

Our analysis provides several hints on the origins of SBOs. The slow streamer swelling occurs on similar time scales as the slow prominence rise, particularly evident of polar crown prominences. Many of the SBOs are related to polar crown eruptions. The SBO locations follow the evolution of quiet sun and polar crown prominences (Figure 5) and the location of the global HCS (Figure 6). The observation of long-duration outflows followed by the streamer disappearance in the wake of the CME is consistent with the reconnection and opening up of a significant portion of the HCS over the source location, which consequently takes time to reclose and reform in a slightly different location. The strong association of SBOs with MFR signatures (Figure 8) further suggests that the entrained structures are extended along the line of sight because such configuration will result in better viewing of the CME structure. The slow evolution, from swelling to CME appearance in the coronagraph FOVs, the generally weak surface signatures, and the frequent flux rope signatures, all point to a slowly energizing ejection process which involves reconnection (albeit high in the corona) across a PIL to release the CME. This description fits quite nicely with the Lynch et al. (2016) model of 'stealth' CMEs (Robbrecht et al. 2009). The model shows that these events are driven by the release of magnetic energy accumulated by differential rotation and, hence, are not intrinsically different that flare-related CMEs except for the height of the reconnection onset and the smaller amounts of magnetic energy available. 'Stealth' CMEs may be just a subset of SBO CMEs defined by the height where the final reconnection (and release) occurs. In other words, we find that SBOs are as magnetically driven as the average CME but SBOs originate in extended PILs, over the quiet sun and are powered by slow injection of magnetic energy, possibly through differential rotation.

We uncover several interesting aspects in the solar cycle variation of SBOs. First of all, our 19 years of observations show that the SBO rate changes during the cycle, in disagreement with the Howard et al. (1986) assertion of a constant SBO rate, which was based on a much smaller and temporally restricted sample. Second, there is a marked change in the SBO properties between SC 23 and 24 (Figure 5). In SC 23, SBOs are faster and more energetic (shorter duration, higher speed and kinetic energy) but less frequent towards the maximum, suggesting that some of those SBOs may be associated with active regions. We see no such trend in SC 24. Instead, the SBO duration keeps increasing through the maximum while the speed and kinetic energy remain constant.

The widths exhibit an even more remarkable solar cycle evolution. There is no obvious correlation with the cycle phase. In SC 23, the widths remain constant from the SC 22/23 minimum through the maximum, peak in 2004, and then decline in tandem to the SC 23 decline. In SC 24, however, the widths increase, almost monotonically until the end of our sample. There is a particularly sharp increase in SBO widths in 2014-2015 which is not reflected in the speeds. We have revisited and repeated many of our measurements during the period to understand whether the sharp increase could be due to measurement errors and find that they are not. Many of our SBO identifications are related to halo or partial halo CMEs, which are the likely reason for the sharp width increase. Many of these wide CMEs in SC 24, however, are not fast, so their larger widths are unlikely to come from shock signatures (e.g. Kwon et al. 2015). The 2014-2015 speeds remain constant but there is a small uptick in the masses and, consequently, in the kinetic energy. Therefore, the widths likely reflect the larger size of SC 24 SBOs compared to their SC 23 counterparts. The obvious implication is that the SC 24 SBOs originate from longer PILs and hence involve the ejection of larger magnetic structures but this remains to be established via a more thorough investigation.

SBO-CMEs may have some implications for Space Weather studies. Given their slow evolution, SBOs can be extremely hard to detect if they are directed towards the observer (e.g. Figure 2 in Robbrecht et al. 2009). They require a wide FOV to ensure that their imprint in the background corona will be captured. If they lack strong surface signatures, which is likely given their gradual nature, SBOs will often be missed by the observers. It is therefore quite likely that Earth-directed SBOs are not accounted for properly in operational models. Any interactions with faster transients will be missed leading to forecasting errors in CME time-of-arrival or their other properties. As far as we know, this issue has not been researched yet and the effect of SBOs on Space Weather (albeit indirect) remains unknown.

## 5. Conclusions

We have presented a comprehensive analysis of a particular class of CMEs, called streamer-blowout (SBOs), marked by a gradual swelling of the streamer before the event and the near (many times, complete) evacuation of the streamer after the event. We identified 909 SBO events in 19 years of LASCO/C2 data (1996-2015) and have extracted physical parameters for a subset (Table 1) from the CDAW CME catalog. Their statistical analysis and comparison with magnetic field observations reveal the following:

- *Physical parameters*: SBOs are more massive and wider than a typical CME. The average SBO-CME has a width of 72°, speed of 390 km/s and mass of $3 \times 10^{15}$ g (Table 1). All physical parameters have an inverse correlation with event duration.
- *Blowout Duration*: The average duration of a blowout CME, from start of streamer swelling to complete evacuation, is 40.5 hours but most blowouts average 26 hours as stated in the text following Table 1. Some events can extend to 4 days or more. The duration has a weak solar cycle dependence (for SC 23 only) with shorter durations during the maximum phase.
- *Occurrence Rates:* SBOs are relatively infrequent during minimum at a rate of 2 per month rising to 6 per month at maximum for SC 23. However, the peak SBO rate has doubled in SC 24.
- *Origins*: SBOs originate only in streamers and their locations follow the global dipolar field. The SBO position angles follow the HCS tilt (Figure 6) although this pattern appears broken during the decline of SC 24.
- *Solar Cycle*: All SBO parameters show some variations with solar cycle but do not correlate with sunspot numbers (Figures 5 and 7). This suggests that they originate from polarity inversion lines (PILs) outside active regions.
- *Morphology*: Flux rope signatures within SBO-CMEs occur at a higher rate than in the general CME population (61% versus 40%). SBO-CMEs may suffer fewer projection effects that could mask the flux rope signatures because they are wider and slower than a typical CME. This suggests that SBO-CMEs likely arise from structures associated with extended PILs (e.g. polar crown filaments, quiet sun filaments) which is consistent with the good correlation between their position angle and the HCS tilt and the lack of correlation with sunspot numbers.

Taken together, these results imply that SBO CMEs originate in quiet sun and polar crown PILs, through the release, via reconnection, of magnetic energy, likely accumulated via differential rotation. They seem to follow the evolution of the global (dipole) magnetic field, rather than higher order multipoles, and, hence, reflect the differences between SC 23 and 24 more prominently than CMEs from active regions. The drivers behind their slow evolution and their role in Space Weather are not yet established. Overall, SBO CMEs are similar to the 'standard' CME but with some important differences.

**Acknowledgements**

A.V. is supported by NASA grant NNX16AH70G and the LWS program though NNX15AT42G under ROSES NNH13ZDA001N. D.F.W is supported by Navy grant N00173-14-1-G014. We thank the referee for the careful reading and useful comments. The SOHO/LASCO data used here are produced by a consortium of the Naval Research Laboratory (USA), Max-Planck-Institut fuer Aeronomie (Germany), Laboratoire d' Astronomie (France), and the University of Birmingham (UK). SOHO is a project of international cooperation between ESA and NASA. The LASCO CME catalog is generated and maintained at the CDAW Data Center by NASA and The Catholic University of America in cooperation with the Naval Research Laboratory.